\documentclass[a4paper,11pt]{article}
\pdfoutput=1 % if your are submitting a pdflatex (i.e. if you have
\usepackage{jcappub} % for details on the use of the package, please
\usepackage{multirow, tabularx}
\usepackage[T1]{fontenc} % if needed

\title{Compact objects in conformal nonlinear electrodynamics}

\author[a]{I.P. Denisova,}
\author[b]{B.D. Garmaev,}
\author[b,1]{V.A.Sokolov \note{Corresponding author.}}

\affiliation[a]{Moscow Aviation Institute (National Research University), Moscow, Russia}
\affiliation[b]{Physics Department, Moscow State University, Moscow, Russia}

\emailAdd{sokolov.sev@inbox.ru}

\abstract{In this paper we consider a special case of vacuum non-linear
electrodynamics with a stress-energy tensor conformal to the
Maxwell theory. Distinctive features 
of this model are: the absence of dimensional parameter for 
non-linearity description and a very simple form of the dominant energy 
condition, which can be easily verified in an arbitrary
pseudo-riemannian space-time with the consequent  
constrains on the model parameters.
In this paper we analyse some properties of astrophysical 
compact objects coupled to conformal vacuum non-linear electrodynamics.}

\keywords{vacuum non-linear electrodynamics, conformal field theory}

\begin{document}
\maketitle
\flushbottom

\section{Introduction}\label{sec:intro}

Electromagnetic field theory suggesting the possibility
of non-linear processes in vacuum due to complicated dependence 
of Lagrangian on both electromagnetic field invariants  
is usually called vacuum non-linear electrodynamics. 
Despite of the extremely encouraging results in observing of
vacuum birefringence in the pulsar strong magnetic field~\cite{a1}  
predicted by some non-linear models,  experimental 
status of vacuum electrodynamics remains to be unrevealed.
The advances in physics of intense electromagnetic field  
and observational X-ray astronomy give a hope 
for new experimental results in several projects 
such as extremely intensive laser facilities like 
ELI~\cite{a2}, XFEL~\cite{a3}, Apollon~\cite{a4}, 
XCELL~\cite{a5} and orbital X-ray polarimeters like 
XIPE~\cite{a6} and IXPE~\cite{a7}.

Early theoretical assumptions
for vacuum electrodynamics non-linearity
were proposed in Born-Infeld~\cite{a8} and
Heisenberg-Euler~\cite{a9} theories. 
Born-Infeld electrodynamics is a phenomenological model based on  
the assertion of finiteness of the electromagnetic field energy for  
charged point-like particle. 
While time later it was shown that this model
arise in the string theory  as an effective  action for abelian 
vector field coupled to virtual open Bose string~\cite{a10}.
As well as in Maxwell electrodynamics, Born-Infeld theory
induces a dual invariance~\cite{a11} and displays
no birefringence in vacuum~\cite{a12}.
Undoubted advantage of this theory is strict
and relatively simple form of the Lagrangian, which opens
a possibilities to find an exact solutions.
Nevertheless there are some disadvantages.
The first one was noted by the authors immediately
after establishing of the theory.
The value of the electric field in the center of the point-like charge
depends on the direction of approach to it.
Resolving this problem leads to the modified Lagrangian~\cite{a13}.
The principle used for Born-Infeld Lagrangian constructing turned out 
to be extremely productive and found application in other  theoretical areas, 
for instance, it was implemented in several  modifications of Born-Infeld gravity~\cite{a14}
and after supplementation by AdS/CFT correspondence it 
was used for a holographic superconductors 
description~\cite{a15,a16}.

Heisenberg-Euler theory~\cite{a9} originates from quantum electrodynamics (QED)
based on Maxwell theory and 
consider the radiative corrections to vacuum polarization
in the external electromagnetic field, which leads to
vacuum behaviour as a continuous media with a non-linear features.
This theory, is seems to be, the most profound and justified, 
especially since some of it's predictions have found experimental 
confirmation in a subcritical or perturbative regime. 
First of all, this refers to the electron anomalous magnetic moment correction,
which still remains an example of unprecedented correspondence between the
theory and the experiment~\cite{a17}.
Some other QED predictions such as Delbr\"{u}ck scattering~\cite{a18},
photon splitting~\cite{a19}, Lamb shift~\cite{a20},
nonlinear Compton scattering~\cite{a21} 
are also well-established. There is no any exact  
expression for Heisenberg-Euler Lagrangian, and only %for which only 
representation in the form of the series 
of the loop corrections is available. 
This significantly complicates the analysis and often makes it 
impossible to obtain an exact solutions. 
Moreover, due to the  correspondence 
between the Maxwell theory and non-linear electrodynamics models, 
all of these models should lead to the Maxwell theory in 
the weak non-linearity regime. In this case 
application of the quantization procedure and subsequent calculation of the 
loop corrections to any Lagrangian should lead for all of the models 
to the Heisenberg-Euler theory in the leading terms of Lagrangian expansion.
Distinctive features of various non-linear 
electrodynamics models after application of the quantization procedure 
will appear as a modification to the Heisenberg-Euler Lagrangian
in terms of high order of smallness.
The effects coupled with such a terms 
became valuable only in a sufficiently non-linear regime,
with strong electromagnetic field comparable to Sauter-Schwinger limit. 
This regime of QED is poorly understood~\cite{a22} and
provides a new window for experimental and theoretical researches.
So confirmations of the Heisenberg-Euler theory 
predictions does not cancel the question
about the choice of the theoretical model for
non-linear electrodynamics in the classical field theory.

For this reason a set of new empirical models for non-linear 
electrodynamics has been proposed. 
Among them, the special place is given to
the models inspired by astrophysics and cosmology~\cite{a23,a24,a25},
the development of which  opened up an unusual theoretical view on the
Universe acceleration due to non-linear electromagnetic
processes~\cite{a26,a27}. Moreover, charged regular black holes as a
new class of compact astrophysical objects were predicted
in~\cite{a28,a29}.
As a rule, the choice of the Lagrangian for the model is heuristic 
and primarily based, on possibility to find an exact
analytical solution for the case under consideration.
At the same time, of great interest is the study of the models grounded on 
the more profound principles, one of which may be postulated as  
the maximal retention of the Maxwell's theory properties and simultaneously 
prediction of the vacuum non-linear response.
In this paper we consider a general class of vacuum non-linear  
electrodynamics with the zero trace of the stress-energy tensor. 
This condition is sufficient for the model to retain all 
group symmetries of the Maxwell theory i.e. invariance under 
Poincar\'{e} group, coordinate scaling and the conformal group~\cite{a30}.
Another distinctive feature of such a models is the lack of a 
dimensional parameter describing the non-linearity. 
For instance, in Born-Infeld model, such parameter 
is the value of the field strength in the 
center of the point-like charge, and in the Heisenberg-Euler 
theory -- this is the characteristic quantum induction.  
However, for the models under consideration in the paper, 
this parameter should be dimensionless,
and probably, it can be expressed as the combination of the 
fundamental constants.
It should be noted, that there are descriptions 
for particular models of non-linear electrodynamics 
with the zero trace of the stress-energy tensor.
The most vivid of them are~\cite{a31} and~\cite{a32}
are devoted to the charged black holes and their thermodynamics.

In this paper, we consider the most general form of the traceless 
non-linear electrodynamics and describe some properties of 
the exact solutions for the compact astrophysical objects in such a model.
This continues the series of papers, started by~\cite{a33}
in which the vacuum birefringence for a general case of the traceless models
was described.

The paper is organized as follows: In section~\ref{S2},
we obtain general form of the traceless non-linear electrodynamics 
Lagrangian and discuss some of it's features. In section~\ref{S3} 
we set fundamental restrictions on the Lagrangian. In section~\ref{S4}
we obtain analogue of the Reissner-Nordstr\"{o}m solution 
of the black hole with the dyon charge. Section~\ref{S5}
is devoted to the analogue of the Vaidya-Bonnor solution and it's features.
In section~~\ref{S6} we describe Kerr-Newmann black hole in conformal 
non-linear electrodynamics, and 
in the last section we summarize our results.
For more convenience we will use geometerized units ($G=c={\hbar}=1$) 
and the metric signature   $\{ +\ , -\ , -\ , -\ \}$. 

\section{Conformal vacuum nonlinear electrodynamics}\label{S2}
Let us consider a general form of the action for Lorentz-invariant 
vacuum non-linear electrodynamics in the space-time with the metric tensor $g_{ik}$:
\begin{equation}\label{Action_NED_Gen}
S_m=\int \sqrt{-g} {\cal L}(J_2, J_4) d^4x,
\end{equation}
where Lagrangian ${\cal L}$ is an arbitrary function of 
the electromagnetic field tensor $F_{ik}$
invariants $J_2=F_{ik}F^{ki}$ and $J_4=F_{ik}F^{kl}F_{lm}F^{mi}$,
and $g$ is the determinant of the metric tensor.
Varying the action by the metric $g^{ik}$ it's easy to derive a
symmetric stress-energy tensor for the action~(\ref{Action_NED_Gen}):
\begin{eqnarray}\label{Str_Energ_Gen}
T_{ik}=\frac{2}{\sqrt{-g}}\frac{\delta (\sqrt{-g}{\cal L})}{\delta g^{ik}}=
4\Big[\frac{\partial {\cal L}}{\partial J_2}+
J_2\frac{\partial {\cal L}}{\partial J_4}\Big]F_{ik}^{(2)}+ 
\Big[(2J_4-J_2^2)\frac{\partial {\cal L}}{\partial J_4}-{\cal L}\Big]g_{ik},
\end{eqnarray}
which trace is
\begin{equation}\label{Str_Energ_Trace}
T=T^i_{i}=4\Big[\frac{\partial {\cal L}}{\partial J_2}J_2+
2J_4\frac{\partial {\cal L}}{\partial J_4}-{\cal L}\Big],
\end{equation}
and for brevity we introduce the notation for the second power of the
electromagnetic field tensor $F^{(2)}_{ik}=F_{im}F^{m\ \cdot}_{\cdot\ \ k}$.
It is well known that the Maxwell theory, which corresponds to the 
particular choice ${\cal L}={J_2}/{16\pi}$, leads to 
the traceless stress-energy tensor. 
To retain this feature for vacuum non-linear
electrodynamics let us consider the models with the action~(\ref{Action_NED_Gen}),
which stress-energy tensor is conformal to the Maxwell electrodynamics:
\begin{equation}\label{Str_Energ_Conf}
T_{ik}=\Omega(J_2, J_4)T_{ik}^M=
\frac{\Omega(J_2, J_4)}{4\pi}\Big\{F_{ik}^{(2)}-\frac{g_{ik}}{4}J_2\Big\}, \quad
\end{equation}
where $\Omega(J_2, J_4)$  is an arbitrary function of electromagnetic
invariants. It is easy to see, that this requirement is fully similar to the 
traceless condition:
\begin{equation}\label{Trace_Cond}
J_2\frac{\partial {\cal L}}{\partial J_2}+
2J_4\frac{\partial {\cal L}}{\partial J_4}-{\cal L}=0.
\end{equation}
The model with  the Lagrangian satisfying 
the equation~({\ref{Trace_Cond}}) we will call traceless or 
conformal non-linear electrodynamics (CNED). 
Such a model name is justified, because the Lagrangians which 
satisfying~(\ref{Trace_Cond}) are turned to be invariant under the group 
of conformal-metric transformations 
$g_{ik}\to \tilde{g}_{ik}=\lambda^2(x)
g_{ik}$, where $\lambda$ is an arbitrary, 
scalar multiplier. 
The similar group symmetry is also inherent 
to the  Maxwell theory.

The CNED Lagrgangians have an another one distinctive feature:
the combination of CNED Lagrangians, 
under the certain condition, can be also a CNED Lagrangian. 
To obtain this condition let us consider the  
function ${\cal L}={\cal L}({\cal L}_1,{\cal L}_2)$
of the Lagrangians ${\cal L}_1$ and ${\cal L}_2$, 
which are a solutions of the traceless equation.
After the substitution of ${\cal L}$ to (\ref{Trace_Cond}) 
\begin{eqnarray}\label{S_aux}
\frac{\partial {\cal L}}{\partial {\cal L}_1}
\Big[J_2\frac{\partial {\cal L}_1}{\partial J_2}+
2J_4\frac{\partial {\cal L}_1}{\partial J_4}\Big]+
\frac{\partial {\cal L}}{\partial {\cal L}_2}
\Big[J_2\frac{\partial {\cal L}_2}{\partial J_2}+
2J_4\frac{\partial {\cal L}_2}{\partial J_4}\Big]-{\cal L}({\cal L}_1,{\cal L}_2)=0,
\end{eqnarray}
and taking into account that ${\cal L}_1$ and $ {\cal L}_2$ 
are also CNED Lagrangians, one get the equation, 
any solution of which will retain conformal features:
\begin{equation}\label{S_main}
{\cal L}_1\frac{\partial {\cal L}}{\partial {\cal L}_1}+
{\cal L}_2\frac{\partial {\cal L}}{\partial {\cal L}_2}-{\cal L}({\cal L}_1,{\cal L}_2)=0.
\end{equation}

The property noted above does not limit the possibilities of CNED Lagrangians constructing. For instance, to provide correspondence to the Maxwell theory
we choose the Lagrangian ${\cal L}_1=J_2/16\pi$. Another one Lagrangian 
can be chosen in the form ${\cal L}_2=W(J_2/\sqrt{2J_4})$, which does not satisfy~(\ref{Trace_Cond}), and consequently it is not 
conformally invariant, where $W$ is an arbitrary 
function. Nevertheless, the production of these Lagrangians will satisfy equation~(\ref{Trace_Cond}),  
and it is easy to verify, that it will represent
the most general form of CNED Lagrangian:
\begin{equation}\label{CNED_L}
{\cal L}={\cal L}_1 {\cal L}_2=\frac{J_2}{16\pi} W\Big(\frac{J_2}{\sqrt{2J_4}}\Big)=
\frac{J_2}{16\pi} W(z),
\end{equation}     
where the invariants ration $z=J_2/\sqrt{2J_4}$ 
varies from $z=-1$ for the pure magnetic field,
to the $z=1$, for the pure electric field. Also it should be noted, 
that $z$ is dimensionless, so there is no any dimensional parameter 
to scale the model nonlinerity. And finally, to obtain 
the Maxwell electrodynamics limit, one should take $W=1$.

Despite on the fact that for the Lagrangian~(\ref{CNED_L})
the traceless condition will be met with an arbitrary function $W$, there are 
some valuable restrictions on this function, coming from  fundamental 
principals. 

\section{Fundamental restrictions}\label{S3}
The choice of the function $W(z)$ for each particular CNED 
model must fulfil  fundamental principles, the primary from
which are unitarity and causality conditions. 
The causality principle guarantee that the group 
velocity for the elementary electromagnetic excitations do
not exceed the speed of light in the vacuum. While 
the unitarity criteria provides the positive
definiteness of the norm of every elementary excitation
of the vacuum. The general constrains on the Lagrangian 
which are necessary for causality and unitariry are 
complicated and extremely difficult to analyse.   
So as the same as in~\cite{a34} we will consider 
more particular case when the electric and magnetic fields 
meet an additional requirement $({\bf E} {\bf B})=0$ in the 
certain Lorentz frame. This corresponds to $z=\pm 1$. 
For described field configuration in paper~\cite{a34}
it was obtained a set of
inequalities for the Lagrangian, which guarantee fulfilment of
causality and unitarity criteria:
\begin{equation}\label{M2_Uni1}
\frac{\partial {\cal L}}{\partial J_2}\geq0,\qquad \frac{\partial
{\cal L}}{\partial J_4}\geq 0,
\qquad
\frac{\partial {\cal L}}{\partial J_2}+J_2\frac{\partial {\cal
L}}{\partial J_4}\geq 0,
\end{equation}
\begin{equation}\label{M3_Cas1}
\frac{\partial {\cal L}}{\partial J_2}+J_2\frac{\partial {\cal
L}}{\partial J_4}+2J_2\Big[\frac{\partial^2{\cal L}}{\partial
J_2^2}+2J_2\frac{\partial^2{\cal L}}{\partial J_2 \partial
J_4}+J_2^2\frac{\partial^2{\cal L}}{\partial J_4^2}+\frac{\partial
{\cal L}}{\partial J_4}\Big]\geq 0,
\end{equation}
\begin{equation}\label{M3_Cas2}
\frac{\partial^2{\cal L}}{\partial
J_2^2}+2J_2\frac{\partial^2{\cal L}}{\partial J_2 \partial
J_4}+J_2^2\frac{\partial^2{\cal L}}{\partial J_4^2}+\frac{\partial
{\cal L}}{\partial J_4}\geq 0.
\end{equation}
By taking into account that in general CNED Lagrangian
have a form~(\ref{CNED_L}), one can significantly simplify  
these inequalities which, finally leads to only two non-trivial
constraints:
\begin{equation}\label{CNED_Cas_Uni}
W(z=\pm 1)\geq0, \qquad W'(z=\pm 1)\leq0, 
\end{equation}
where the prime denotes the derivative with the respect to $z$.  

This restriction is not only because 
the Lagrangian must be agreed with the energy
conditions. The class of conformal electrodynamics is
special, because it have very simple restrictions
following from the dominant energy condition~\cite{a35}.
This condition claims that every time-like observer
will find field energy density to be non-negative,
and energy flux to be a causal vector (time-like or null).
These requirements ensures dominance of the energy density
over the other components in the stress-energy tensor.
The dominant energy condition leads to the inequalities:
\begin{equation}\label{Domonant_cond1}
T_{ik}a^i a^k \geq0, \qquad T_{ki}T^{im}a_m a^k \geq 0,
\end{equation} 
where $a^k$ is any arbitrary time-like  vector pointing to the 
future. In general, after implementation to an arbitrary 
Lorentz-invariant non-linear electrodynamics~(\ref{Action_NED_Gen}) the 
inequalities takes a  slightly cumbersome form:
\begin{equation}\label{Gen_NL_DEC1}
T_{ik}a^ia^k=4\Big\{\frac{\partial {\cal L}}{\partial J_2}+
J_2\frac{\partial {\cal L}}{\partial J_4}\Big\}F_{ik}^{(2)}a^ia^k+ 
\Big\{(2J_4-J_2^2)\frac{\partial {\cal L}}{\partial J_4}-{\cal L}\Big\}a_ka^k\geq0,
\end{equation}
\begin{equation}\label{Gen_NL_DEC2}
T_{ki}T^{im}a_m a^k=8\Big\{\frac{\partial {\cal L}}{\partial J_2}+
J_2\frac{\partial {\cal L}}{\partial J_4}\Big\}\times 
\Big\{J_2\frac{\partial {\cal L}}{\partial J_2}+
2J_4\frac{\partial {\cal L}}{\partial J_4}-{\cal L}\Big\}F_{ik}^{(2)}a^ia^k 
\end{equation}
\begin{equation*}
+\Big\{2\Big[\frac{\partial {\cal L}}{\partial J_2}+
J_2\frac{\partial {\cal L}}{\partial J_4}\Big]^2(2J_4-J_2^2)+
\Big[(2J_4-J_2^2)\frac{\partial {\cal L}}
{\partial J_4}-{\cal L}\Big]^2\Big\}a_ka^k\geq 0,
\end{equation*} 
however they become much simpler in the context of CNED with the Lagrangian~(\ref{CNED_L}):
\begin{equation}\label{NL_DEC_CONF1}
T_{ik}a^ia^k=\frac{1}{4\pi}\Big[W+z(1-z^2)W'\Big]
\times \Big\{F_{ik}^{(2)}a^ia^k-\frac{J_2}{4}a_ka^k\Big\}\geq0,
\end{equation}
\begin{equation}
T_{ki}T^{im}a_m a^k=\frac{J_4}{64\pi^2}\Big[W+z(1-z^2)W'\Big]^2\times
\Big\{1-\frac{z^2}{2}\Big\}a_ka^k\geq0.
\end{equation}
The last inequality is always satisfied due to $|z|\leq 1$ and $J_4 \geq 0$,
while the last multiplier in~(\ref{NL_DEC_CONF1}) represents the 
energy condition for Maxwell theory and it is not negative. So the dominant 
energy condition for CNED will take place when: 
\begin{equation}\label{CNED_DEC_Rest}
W(z)+z(1-z^2)W'(z) \geq0.
\end{equation}
This inequality corresponds to the first one in~(\ref{CNED_Cas_Uni})
for pure electric or magnetic field, when $z=\pm 1$. 
The restrictions obtained  in the form of 
inequalities~(\ref{CNED_Cas_Uni}) and~(\ref{CNED_DEC_Rest})
allows to find some constrains on the model parameters and they will be 
employed in the following sections for the analysis of 
consistency with the fundamental principles.

After discussion the general properties and the model
restrictions we proceed to description of an exact solutions for the compact 
objects in CNED.

\section{Reissner-Nordstr\"{o}m black hole}\label{S4}
Primarily we describe solution for the stationary black hole, with  electric and magnetic charges in Einstein gravity with non-zero cosmological constant $\Lambda$.The action functional in this case can be represented in the form:
\begin{equation}
S=-\int\frac{R-2\Lambda}{16\pi}\sqrt{-g} \ d^4x+	
\int{\cal L}\sqrt{-g} d^4x, 
\end{equation}
where $R$ is the scalar curvature, and ${\cal L}$ is the 
Lagrangian density (\ref{CNED_L}).
By varying the action, it is easy to derive electromagnetic field
and Einstein equations:
\begin{eqnarray}
R_{ik}-\frac{R}{2}g_{ik} +\Lambda g_{ik}&=&8\pi T_{ik}, \label{Einsten_eq} \\
\frac{1}{\sqrt{-g}}\frac{\partial \sqrt{-g} Q^{kn}}{\partial x^n}&=&-{4\pi j^k_{(e)}}, 
\label{CNED_Em_Field_eq} \quad
\frac{1}{\sqrt{-g}}\frac{\partial \sqrt{-g} *F^{kn}}{\partial x^n}=-{4\pi j^k_{(m)}},
\end{eqnarray}
where $T_{ik}$ is CNED stress-energy tensor~(\ref{Str_Energ_Conf}), 
$*F^{kn}=e^{knlm}F_{lm}/2\sqrt{-g}$
is dual conjugated electromagnetic field tensor, 
$j^k_{(e)}$ and $j^k_{(m)}$ are current density vectors for electric 
and magnetic charges, and auxiliary tensor $Q^{kn}$ 
can be expressed in the form:
\begin{equation}\label{Q_tens}
Q^{kn}=WF^{kn}+z\Big(F^{kn}-\frac{J_2}{J_4}F_{(3)}^{kn}\Big)W',
\end{equation} 
where the prime denotes the derivative of $W$ with the respect of it's 
argument and $F^{kn}_{(3)}=F^{kl}F_{lm}F^{mn}$ 
is the third power of the field strength 
tensor. We consider the line element of the static spherically 
symmetric space-time:
\begin{equation}\label{Line_elem}
ds^2=e^{2\alpha(r)}dt^2-e^{2\beta(r)}dr^2-r^2(d\theta^2+\sin^2\theta d\varphi^2),
\end{equation} 
and we also suppose the most general form for the field strength tensor
for static point-like charge:
\begin{equation}\label{F_tens}
F_{ik}=E(r)\{\delta_i^0\delta_k^1-\delta_i^1\delta_k^0\}-
B(r)r^2\sin\theta\{\delta_i^2\delta_k^3-\delta_i^3\delta_k^2\},
\end{equation}
where $E(r)$ and $B(r)$ are radial electric and magnetic fields.
Under the chosen symmetries the invariants of the electromagnetic field 
and the dimensionless parameter $z=J_2/\sqrt{2J_4}$ take a form:
\begin{equation}\label{Invar_and_Z}
J_2=2[e^{-2(\alpha+\beta)}E^2-B^2],\quad J_4=2[e^{-4(\alpha+\beta)}E^4+B^4],
\quad z=\frac{e^{-2(\alpha+\beta)}E^2-B^2}{[e^{-4(\alpha+\beta)}E^4+B^4]^{1/2}}.
\end{equation}    
As the point-like source is located at the coordinate center,  non 
zero components of the current densities are:
\begin{equation}
j^{0}_{(e)}=\frac{Q_c}{4\pi r^2}e^{-(\alpha+\beta)}\delta(r), \qquad 
j^{0}_{(m)}=\frac{Q_t}{4\pi r^2}\delta(r),
\end{equation}
where $Q_c$ is the electric charge and $Q_t$ is 
the topological charge of magnetic monopole.
To obtain solutions for the electromagnetic field equations 
we consider the ansatz:
\begin{equation}\label{E_and_B}
E(r)=Q_e e^{\alpha+\beta}/r^2, \qquad B(r)=Q_t/r^2,
\end{equation}
where $Q_e$ is an integration constant coupled with the electric 
and magnetic charge of the black hole.
As the field strength components have similar dependence on coordinates, 
the argument $z$ takes a constant value and it can vary from 
$z=1$ for the pure electric field, to $z=-1$ for the pure magnetic field.
By using the auxiliary expressions~(\ref{Invar_and_Z}) and~(\ref{Q_tens}),
it is easy to find that~(\ref{E_and_B})
is the solution of the electromagnetic field equations under 
the condition that the integration constant 
$Q_e$ and the charges $Q_c$ and $Q_t$ are coupled by the relation:
\begin{equation}\label{Qe_Qc_relation}
Q_e\Big\{W(z)+\frac{z(1-z^2)}{2}\Big[1+
\frac{1-z^2}{1+\mbox{sgn}(z)\sqrt{1-(1-z^2)^2}}\Big]W'(z)\Big\}=Q_c, 
\quad \mbox{where} \quad 
z=\frac{Q_e^2-Q_t^2}{(Q_e^4+Q_t^4)^{1/2}}.
\end{equation}
Similar relation for the charges was obtained earlier
in~\cite{a32} for a  particular choice of the 
electromagnetic field Lagrangian. Also it was obtained by the 
authors, that for the chosen model
$Q_e<Q_c$ when $Q_t\neq 0$. The appearance of two constants with 
a charge dimension in the description of the point source field can 
be collate with the possible difference between the inertial and 
gravitational masses for a point-like particle.
As the constant $Q_e$ is contained in the expression of the electric 
field strength, it can be called "force-charge" and in contra to 
the constant $Q_c$ which is the multiplier in the source density, 
so it can be called "source-charge" or Coulomb charge. 
At the same time, it is interesting to find condition 
under which the source-charge and the force-charge coincide 
$Q_e=Q_c$ in presence of topological charge $Q_t\neq 0$.  
The claim is satisfied when 
the expression in the curly brackets in~(\ref{Qe_Qc_relation}) 
is equal to unity, which can be done 
by choosing 
\begin{equation}\label{CNED_sp_W}
W(z)=1+\frac{c_1}{z}\sqrt{1-\mbox{sgn}(z)\sqrt{1-(1-z^2)^2}},
\end{equation} where $c_1$ is an arbitrary
dimensionless constant. 
It is easy to verify that the restrictions from 
the causality and unitarity~(\ref{CNED_Cas_Uni}) 
now can be expressed in the form 
$W'(z=\pm 1)=-\sqrt{2}c_1\leq 0$, so the constant $c_1\geq0$
should be positive. At the same time, the dominant 
energy condition~(\ref{CNED_DEC_Rest}) leads to the
inequality:
\begin{equation}\label{Dec_spec_CNED}
W+z(1-z^2)W'=1-
\frac{c_1 z \sqrt{1-\mbox{sgn}(z)\sqrt{1-(1-z^2)^2}}}
{\mbox{sgn}(z)\sqrt{1-(1-z^2)^2}}\geq 0,
\end{equation}
the second term in  which is finite, take
minimal value $-c_1\sqrt{2}$ at $z=-1$ and increases monotonically up to the zero at $z=1$.
%In this case one can obtain that $W+z(1-z^2)W'\to 1-c_1 \hbox{sgn}(z)/\sqrt{2}$.
So the dominant energy condition for described CNED model 
will be satisfied when $0\leq c_1\leq\sqrt{2}$. 
From experimental data for the vacuum non-linear electrodynamics effects it seems that  
nonlinerity is a small correction to the	
Maxwell electrodynamics, so $c_1\ll 1$ and implementation of 
inequality~(\ref{Dec_spec_CNED}) is reliably ensured.	 

The another distinctive case of CNED 
take place when $Q_c=0$ and $Q_e\neq 0$, which means that 
the topological charge $Q_c$ will be the source 
both for the electric and magnetic fields. 
In this case the Lagrangian should be expressed with   
$W(z)=c_1\sqrt{1-\mbox{sgn}(z)\sqrt{1-(1-z^2)^2}}/z$. 
However, from the previous 
consideration, it is obvious that this model
contradicts causality and unitarity conditions~(\ref{CNED_Cas_Uni}) and unlikely 
to be related to the real world.  

We proceed to solution of Einstein equations~(\ref{Einsten_eq}), which 
for static space-time with the line-element~(\ref{Line_elem}) 
have only two non-trivial and independent equations:
\begin{eqnarray}\label{Einstein_eq_final_form}
e^{-2\beta}\Big[\frac{2\beta'}{r}-\frac{1}{r^2}\Big]+\frac{1}{r^2}+
\Lambda=8\pi T_{0 \cdot}^{\cdot 0}, \quad
-e^{-2\beta}\Big[\frac{2\alpha'}{r}+\frac{1}{r^2}\Big]+\frac{1}{r^2}+
\Lambda=8\pi T_{1 \cdot}^{\cdot 1},
\end{eqnarray}  
where non-zero components of the stress-energy tensor for
the chosen electromagnetic field configuration are: 
\begin{equation}\label{CNED_Str-En_comp}
T_{0 \cdot}^{\cdot 0}=T_{1 \cdot}^{\cdot 1}=-T_{2 \cdot}^{\cdot 2}=-T_{3 \cdot}^{\cdot 3}
=\frac{Q_e^2+Q_t^2}{8\pi r^4}\Big[W(z)+z(1-z^2)W'(z)\Big],
\end{equation}
and the prime denotes the derivative
with the respect to the correspondent argument.
Subtracting the second equation~(\ref{Einstein_eq_final_form}) from the first one, we obtain the 
condition $\alpha+\beta=f(t)$, where arbitrary function $f$ can be taken equal to zero 
after the choice of the time scale, so 
the solution for both equations can be written in the following form:
\begin{equation}\label{RN_metr_sol}
g_{00}(r)=e^{2\alpha(r)}=e^{-2\beta(r)}=1-\frac{2M}{r}+\frac{{\cal K}}{r^2}+\frac{1}{3}\Lambda r^2,
\end{equation}
where the integration constants were chosen to obtain 
asymptotic limit to the solution in Einstein-Maxwell
theory, so $M$ is the black hole mass and for 
brevity, as in~\cite{a32}, we use the notation:
\begin{equation}\label{K}
{\cal K}= \{Q_e^2+Q_t^2\}\Big[W(z)+z(1-z^2)W'(z)\Big]. 
\end{equation}
Obtained solution corresponds to the Reissiner-Nordstr\"{o}m black hole in CNED and differs from the similar one obtained in~\cite{a32} by more general form 
of the parameter ${\cal K}$ coupled to an 
arbitrary CNED Lagrangian, so the analysis of the black hole thermodynamics
performed in~\cite{a32}  can be completely applied to~(\ref{RN_metr_sol}).    
In this paper the authors distinguish thee different classes of the black hole.
The first one corresponds to the case when ${\cal K}>\Lambda/12$ and was called 
{\it fast black holes}. Phase transitions are absent for this black hole configuration.
For the second class, called {\it slow black holes},  $0<{\cal K}<\Lambda/12$ 
and  there are two phase transitions.
These two types of black holes have their analogues in Einstein-Maxwell theory.
The third class {\it inverse black holes} corresponds to ${\cal K}<0$ and 
posses solely one phase transition. This type of black hole is typical for 
conformal-invariant electrodynamics, and in the more
special case for inverse electrodynamics 
model proposed in~\cite{a32}. However,
it should be noted that the authors 
did not consider fundamental restrictions on the model parameters. 
In particular, the existence of the inverse black holes 
class contradicts to the dominant energy 
condition~(\ref{CNED_DEC_Rest}), the consequence of which is the restriction ${\cal K}\geq 0$.
A global violation of this condition makes possibility 
of such black holes very uncertain.

Let us turn to the  other exact solution for the compact astrophysical object in 
conformal non-linear electrodynamics. 

\section{Vaidya-Bonnor radiating solution}\label{S5}
\begin{figure}[tbp]
\centering
\includegraphics[width=.65\textwidth, clip]{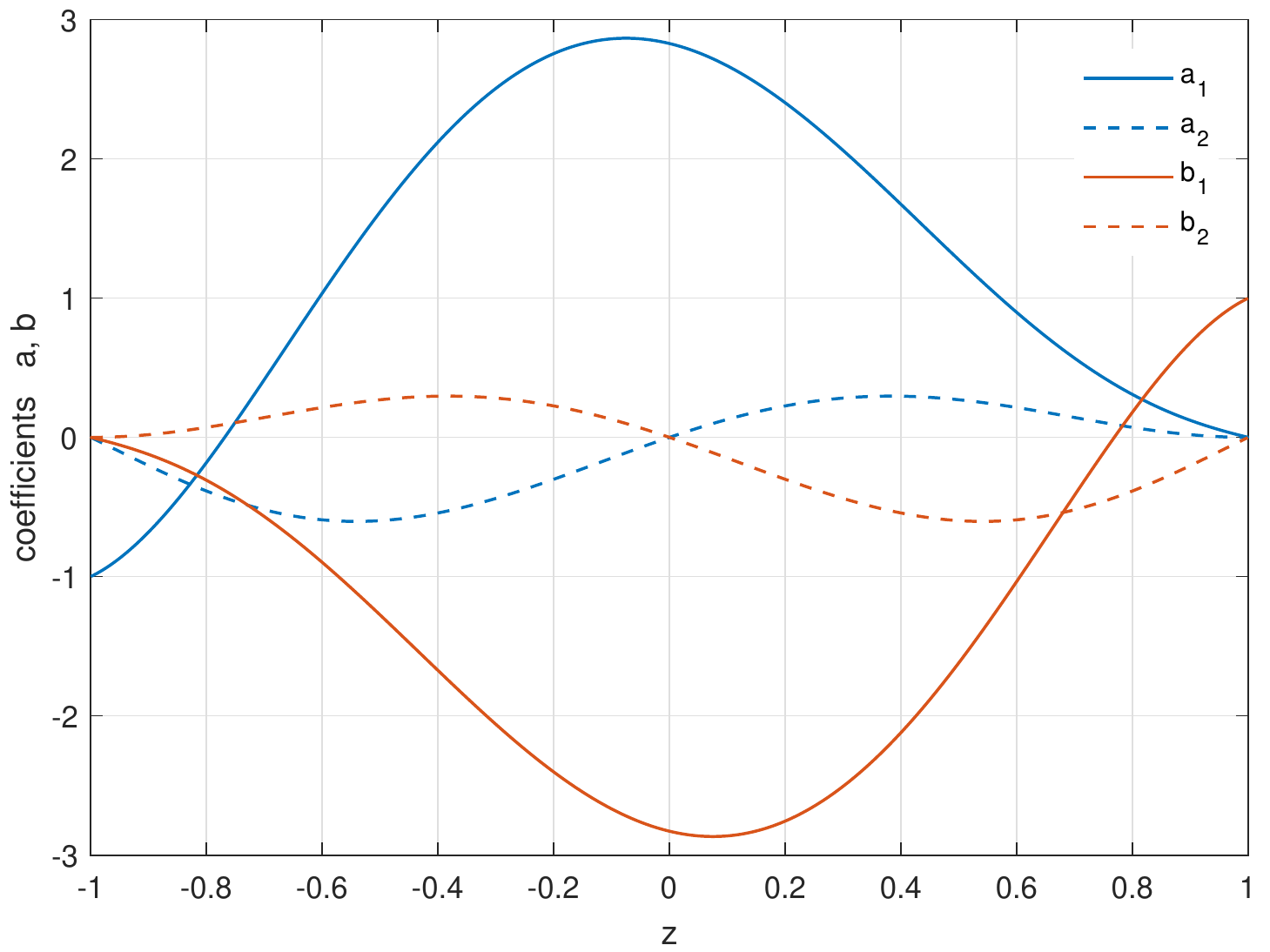}
\caption{ Coefficients  $a$ and $b$ as the functions of $z$.}
\label{fig:1}
\end{figure}

Let us consider solution of the Einstein-CNED equations 
describing the emission of charged null fluid form the
spherically symmetric star with the electric and magnetic charges.  
This solution will be an extension of Vaidya-Bonnor~\cite{a36} metric to 
an arbitrary type of conformal non-linear electrodynamics.   
To obtain  the solution, as traditionally, we suppose the line-element 
in Eddington-Finkelstain coordinates in the form:
\begin{equation}\label{Line_elem_EF}
ds^2=G(u,r)du^2+2du dr -r^2(d\theta^2+\sin^2\theta d\varphi^2),
\end{equation}
where $G$ is the metric function and $u$ is the retarded time.
This form of the line-element can be coupled to~(\ref{Line_elem}) 
buy  the local time transformation $dt=G du+dr/G$,
with additional conditions $e^{2\alpha(t,r)}=1$ and $e^{-2\beta(t,r)}=1/G^2(u,r)$.
It is easy to write non-zero covariant and contravariant metric components:
\begin{equation}
g_{00}=G(u,r), \quad g_{01}=1, \quad g_{22}=-r^2, \quad g_{33}=-r^2\sin^2\theta,
\end{equation}
\begin{equation*}
g^{11}=-g_{00}, \quad g^{01}=1, \quad g^{22}=1/g_{22}, \quad g^{33}=1/g_{33},
\end{equation*}
from which follows that $\sqrt{-g}=r^2\sin\theta$.
As it was earlier in Sec.~\ref{S4}, we represent electromagnetic field 
tensor in the form~(\ref{F_tens}) and assume the 
electromagnetic  field strength as: $B=Q_t/r^2$, $E=Q_e/r^2$.
However, unlike the Reissiner-Nordstr\"{o}m solution the 
topological charge $Q_t=Q_t(u)$, the source-charge $Q_c=Q_c(u)$,
the force-charge $Q_e=Q_e(u)$ and the star mass $M=M(u)$ are any arbitrary functions of 
the retarded time. It is easy to verify that the electromagnetic field satisfies 
equations~(\ref{CNED_Em_Field_eq}) with the same form of the $Q^{kn}$ tensor~(\ref{Q_tens})
and the the following form of the current densities for electric and magnetic charges:
\begin{equation}
J^k_{(e)}=\frac{1}{4\pi r^2}\Big[Q_c\delta(r)\delta_0^k-\dot{Q_c}(u)\delta^k_1\Big], \qquad 
J^k_{(m)}=\frac{1}{4\pi r^2}\Big[Q_t\delta(r)\delta_0^k-\dot{Q_t}(u)\delta^k_1\Big],
\end{equation}   
where the dot denotes the derivative with the respect of the retarded time,
$\delta_i^k$ is the Kronecker symbol and the the source-charge $Q_c(u)$ is coupled to the 
force-charge $Q_e(u)$ by the same relation as earlier in~(\ref{Qe_Qc_relation}). 
To find modified Vaidya-Bonnor metric it is necessary to include the stress-energy 
tensor of the null-fluid in the Einstein equations:
\begin{eqnarray}
R_{ik}-\frac{R}{2}g_{ik} +\Lambda g_{ik}&=&8\pi (T_{ik}+V_iV_k),  
\end{eqnarray}
where $V_i$ is the null fluid current vector, for which $g_{ik}V^iV^k=0$. 
And finally, for  more completeness we also assume the 
Lambda-term to be a function varying with the retarded time $\Lambda=\Lambda(u)$. 
Einstein's equations in this case take a form:
\begin{equation}\label{Einstaen_eq_for_Vaidya-Bonnor}
\frac{ \partial }{\partial r}[r G(u,r)]=1+r^2\Lambda(u)-8\pi r^2(T_{0\cdot}^{\cdot 0}+V_0V^0), 
\qquad
\frac{1}{r}\frac{\partial G (u,r)}{\partial u}=8\pi V_0V^1, 
\end{equation}
where the electromagnetic field stress-energy tensor components 
can be expressed from~(\ref{CNED_Str-En_comp}). As in original paper by 
Vaidya and Bonnor~\cite{a36} we assume the 
null fluid current density to be radial and represent 
it with the scalar $N=N(u,r)$ in the form: $V^i=N\delta^i_1$, 
so the solution of~(\ref{Einstaen_eq_for_Vaidya-Bonnor}) reads as:
\begin{equation}\label{CNED_metr_VB}
G(u,r)=1-\frac{2M(u)}{r}+\frac{{\cal K}(u)}{r^2}+\frac{1}{3}\Lambda(u)r^2, 
\qquad 
N^2=\frac{3\dot{{\cal K}}-6\dot{M}r+\dot{\Lambda}r^4}{24\pi r^3},
\end{equation}
where the expression for ${\cal K}$ coincides with~(\ref{K}) 
and it's derivative is:
\begin{equation}\label{K_dot}
\dot{{\cal K}}=\frac{d}{du}\Big\{\Big[Q_e^2(u)+Q_t^2(u)\Big]\times
\Big[W(z)+z(1-z^2)W'(z)\Big]\Big\}
\end{equation}
\begin{equation*}
=2Q_e\dot{Q}_e\Big\{W+a_1W'+a_2W''\Big\}
+2Q_t\dot{Q}_t\Big\{W+b_1W'+b_2W''\Big\}, 
\end{equation*}
with the auxiliary notations for the coefficients, introduced for brevity:
\begin{equation}
a_1=(1-z^2)\Big[z+
\frac{(2-z^2)(2-3z^2)}{\sqrt{2+2\mbox{sgn}(z)
\sqrt{1-(1-z^2)^2}}}\Big],
\quad a_2=\frac{z(2-z^2)(1-z^2)^2}{\sqrt{2+2\mbox{sgn}(z)
\sqrt{1-(1-z^2)^2}}},
\end{equation}
\begin{equation*}
b_1=(1-z^2)\Big[z-
\frac{(2-z^2)(2-3z^2)}{\sqrt{2-2\mbox{sgn}(z)
\sqrt{1-(1-z^2)^2}}}\Big],
\quad 
b_2=-\frac{z(2-z^2)(1-z^2)^2}{\sqrt{2-2\mbox{sgn}(z)
\sqrt{1-(1-z^2)^2}}}.
\end{equation*}
The dependence of the coefficients $a$ and $b$ on  $z$ 
is represented on the figure~\ref{fig:1}. It should be noted that 
$a_2=0$, $b_2=0$ for a pure magnetic and electric field when $z=\pm 1$. 
These coefficients are also equal to zero, when $z=0$ at $Q_e=\pm Q_t$. 
For the listed cases only the values of $W$ and  $W'$, which are 
restricted by the causality and unitarity conditions~(\ref{CNED_Cas_Uni}),	 
will handle $\dot{\cal K}$ the expressions  
for which in these particular cases can be found 
in the table~\ref{tab:1}.
\begin{table}[h]
	\begin{center}
	  \begin{tabular}{|c|c|c|}
  			\hline 
  		 	& $a$, $b$ & $\dot{\cal K}$ \\[5pt]
  		 	\hline
          	$z=-1$ &\hbox{ $a_1=-1$,  $b_1=0$, $a_2=0$, $b_2=0$} 
          	& $\dot{\cal K}=2Q_t\dot{Q}_t(W-W')$\\[5pt]
  		 	\hline
  		  	\multirow{2}{*}{$z=0$} 
  		  	&\multirow{2}{*}{\hbox{$a_1=2\sqrt{2}$, $b_1=-2\sqrt{2}$, $a_2=0$, $b_2=0$}} 
  		  	&$\dot{\cal K}=2Q_e[\dot{Q}_e(W+2\sqrt{2}W')$ \\
  		  	& & $+\dot{Q}_t(W-2\sqrt{2}W')]$ \\[5pt]
            \hline
  		  	$z=1$ & \hbox{$a_1=0$, $b_1=1$, $a_2=0$, $b_2=0$} & 
  		  	$\dot{\cal K}=2Q_e\dot{Q}_e(W+W')$ \\[5pt]
  		    \hline
	\end{tabular}
\end{center}
\caption{Particular expressions for $\dot{\cal K}$ for some spacial $z$}
\label{tab:1}
\end{table}

To ensure that $N^2$ is not negative for an arbitrary 
distance to the star center, it is necessary to demand that 
the star mass decrease with the retarded time $\dot{M} \leq 0$, 
%as well as  and $\dot{\cal K} \geq 0$, 
the Lambda-term should vary with $\dot{\Lambda} \geq 0$, 
following by the cosmological dynamics, while the 
${\cal K}$ should increase due to ionization processes in the star 
and radiation of the charged fluid.
By the fact, the condition $N^2\geq 0$ should be verified for each particular instance
of the model function $W$, however as it follows from~(\ref{K_dot})
it will be certainly fulfilled when
\begin{equation}
\frac{d}{du}\Big[W(z)+z(1-z^2)W'(z)\Big]\geq 0.
\end{equation}
Moreover, as the coefficients $a$ and $b$ are finite, 
the fulfilment of this condition also should be expected  
in the case when the 
derivatives $W'(z)$ and $W''(z)$ are bounded and 
proportional to a small parameter which describes the model nonlinearity.

In general, the form of the obtained solution for the radiating star~(\ref{CNED_metr_VB})
coincides with the original Vaidya-Bonnor metric and differs 
from it by redefinition of the charge term:
$Q_e^2+Q_t^2\to {\cal K}$.  This fact allows to apply
the features of Vaidya-Bonnor solution to
it's CNED extension~(\ref{CNED_metr_VB}).  
For instance, to describe  thermodynamics of the 
star one can fully use
the results obtained in~\cite{a37, a38}, so the Hawking temperature 
on the event or on the cosmic horizon $r_h$ reads as:  
\begin{equation}
T_H=\frac{r_h-M-\frac{2}{3}\Lambda r_h^2-2r_h\dot{r}_h}
{2\pi k(2M r_h-{\cal K}-\frac{1}{3}\Lambda r_h^4)}, 
\end{equation}
where $k$ is Boltzmann constant and the horizon radius 
can be expressed from the null surface equation:
\begin{equation}
r_h^2-2Mr_h+{\cal K}-\frac{1}{3}\Lambda r_h^4 -2r_h\dot{r}_h=0.
\end{equation}
Using the noted correspondence between the known classical solutions 
of Einstein-Maxwell's equations and their CNED extensions, in the next 
section we will briefly describe one more solution for a compact object 
in the context of a rotating charged black hole. 

\section{Kerr-Newmann black hole}\label{S6}
Finally we consider CNED extension for the rotating black hole 
with the mass $M$, angular momentum per mass unit $a$, 
charged with the electric and magnetic charges $Q_e$ and $Q_t$.
In the absence of rotation, the space-time of such a black hole
is described by the analogue of  
Reissner-Nordstr\"{o}m metric~(\ref{RN_metr_sol}) with the zero 
cosmological constant $\Lambda=0$. 
To derive the solution in the rotational case
it is convenient to apply Newman-Janis algorithm~\cite{a39}
which converts the static spherically symmetric 
space-time in to the rotational one.

The algorithm assumes several steps, and 
at the first one we need to rewrite the line element of the 
static solution with the Eddington-Filkestein coordinates
as it was done in~(\ref{Line_elem_EF}),	 
%\begin{equation}\label{Static_RN_line_elem}
%ds^2=f(r)dt^2+2dt dr -r^2(d\theta^2+\sin^2\theta d\varphi^2),
%\end{equation}
where now $G(r)=1-{2M}/{r}+{{\cal K}}/{r^2}$ 
corresponds to Reissner-Nordstr\"{o}m metric obtained 
earlier in the section~\ref{S4}, and 
the expression for ${\cal K}$ is given by~(\ref{K}).
At the same step it is also necessary to introduce four isotropic vectors 
of the Newman-Penrose tetrade $l^k$, $n^k$, $m^k$, $\bar{m}^k$
and represent the space-time contravariant metric 
tensor as dyad production of these vectors:
\begin{equation}\label{Metr_tetrade}
g^{ik}=l^i n^k+ l^k n^i-m^i \bar{m}^k-m^k \bar{m}^i,
\end{equation} 
where the bar here and below means the complex conjugation.
It easy to verify, that for the Reissner-Nordstr\"{o}m space-time
the components of the tetrade vectors should be taken in form:
\begin{equation}\label{Init_tetrade}
l^k=\delta^k_1, \qquad n^k=\delta^k_0+\frac{1}{2}G(r)\delta^k_1, \qquad
m^k=\frac{1}{\sqrt{2}r}\Big[\delta^k_2+\frac{i}{\sin\theta}\delta^k_3\Big].
\end{equation}	
The second step implies the coordinate mapping to the complex plane 
$r\in \mathbb{R} \to r\in \mathbb{C}$,  $u\in\mathbb{R} \to u\in \mathbb{C}$
with subsequent transformation to the new real coordinates 
set $\{\tilde{u}, \tilde{r}, \tilde{\theta}, \tilde{\varphi}\}$:
\begin{equation}
r\to \tilde{r}=r+ia \cos\theta, \qquad u\to \tilde{u}=u-ia\cos\theta, \qquad
\tilde{\theta}\to\theta, \qquad \tilde{\varphi}\to\varphi.
\end{equation}
This coordinate transformation 
is chosen to retain $l^k$, $n^k$ 
to be real and $m^k$, $\bar{m}^k$
to be complex conjugated to each other, 
even after the transformation. 
So in the new coordinates, the components of the tetrade 
vectors will take a form:
\begin{equation}\label{Transf_tetrade}
l^k\to \tilde{l}^k=\delta^k_0, \quad 
n^k\to \tilde{n}^k=\delta_0^k+\frac{1}{2}\tilde{G}(r, \bar{r})\delta_1^k,
\end{equation}
\begin{equation*}
m^k\to \tilde{m}^k=\frac{1}{\sqrt{2}(\tilde{r}+ia\cos\theta)}
\Big[ia\sin\theta(\delta_0^k-\delta_1^k)+\delta^k_2+\frac{i}{\sin\theta}\delta^k_3\Big].
\end{equation*}
At the same time, to retain the metric function $\tilde{G}$ to be real
the mapping rule for it should be taken in the form   
\begin{equation}\label{G_map}
G(r)\to \tilde{G}(r, \bar{r})=1-M\Big(\frac{1}{r}+\frac{1}{\bar{r}}\Big)+\frac{\cal K}{r\bar{r}}
=1-\frac{2M\tilde{r}-{\cal K}}{\Sigma}, 
\end{equation}
where, as it usual for the Kerr metric, it is convenient to use 
the notation $\Sigma=\tilde{r}^2+a^2\cos^2\theta$. 

The next step in the Janis-Newman algorithm assumes 
construction of the metric tensor according to the expression~(\ref{Metr_tetrade}),
but with use of the transformed vectors~(\ref{Transf_tetrade})
instead of~(\ref{Init_tetrade}).
After the tetrade components substitution we will obtain the contravariant metric tensor 
correspondent to the Kerr-Newman space-time:
\begin{equation}
\renewcommand\arraystretch{1.5}
  \tilde{g}^{ik}=\begin{pmatrix}
    -\dfrac{a^2\sin^2\theta}{\Sigma} &  \ \ \ 1+\dfrac{a^2\sin^2\theta}{\Sigma} & 0 & -\dfrac{a}{\Sigma} \\
    1+\dfrac{a^2\sin^2\theta}{\Sigma} & \ \ \ -\tilde{G}-\dfrac{a^2\sin^2\theta}{\Sigma} & 0 & \dfrac{a}{\Sigma}\\
    0 & 0 & -\dfrac{1}{\Sigma} & 0 \\
    -\dfrac{a}{\Sigma} & \dfrac{a}{\Sigma} & 0 & -\dfrac{1}{\Sigma \sin^2\theta}
  \end{pmatrix}. 
  \label{Meric_KN_Cont}
\end{equation}

At the last step of the algorithm 
one should perform transformation to the Boyer-Lindquist coordinates
$d\tilde{u}=dt+U(\tilde{r}, \theta)d\tilde{r}$, \ 
$d\varphi=d\varphi+\Phi(\tilde{r}, \theta)d\tilde{r}$, where 
\begin{equation}
U(\tilde{r}, \theta)=-\frac{\Sigma+a^2\sin^2\theta}{\tilde{G}\Sigma+a^2\sin^2\theta}, 
\qquad \Phi(\tilde{r}, \theta)=-\frac{a}{\tilde{G}\Sigma+a^2\sin^2\theta}.
\end{equation}
This transformation finally leads to the line element in the form:
\begin{equation}
d\tilde{s}^2=\tilde{G}dt^2-\frac{\Sigma}{\tilde{G}\Sigma+a^2\sin^2\theta}d\tilde{r}^2+
2a\sin^2\theta(1-\tilde{G})dt d\varphi
\end{equation}
\begin{equation*}
-\Sigma d\theta^2-\sin^2\theta\big(\Sigma+a^2(2-\tilde{G})\sin^2\theta\big)d\varphi^2.		
\end{equation*}  
In the following, the tilde should be omitted from the notations,
as well as should be taken that $r\geq 0$. 
A more detailed description and justification of the Janis-Newman 
algorithm can be found in the original paper~\cite{a39}
and in the papers devoted to applications and generalization 
of the method~\citep{a40, a41, a42}.   

Substitution of the metric function~(\ref{G_map}) with account of~(\ref{K}) gives 
explicit form for the line element of rotating charged black hole in 
an arbitrary CNED model:
\begin{equation}\label{KN_line_element_CNED}
ds^2=\Big(1-\frac{2Mr-(Q_e^2+Q_t^2)[W(z)+z(1-z^2)W'(z)]}
{r^2+a^2\cos^2\theta}\Big)dt^2
\end{equation}
\begin{equation*}
-\Big(\frac{r^2+a^2\cos^2\theta}{r^2+a^2-2Mr+(Q_e^2+Q_t^2)[W(z)+z(1-z^2)W'(z)]}\Big)dr^2-(r^2+a^2\cos^2\theta) 
d\theta^2
\end{equation*}
\begin{equation*}
-\sin^2\theta\Big(r^2+a^2+a^2\Big[\frac{2Mr-(Q_e^2+Q_t^2)[W(z)+z(1-z^2)W'(z)]}
{r^2+a^2\cos^2\theta}\Big]
\sin^2\theta\Big)d \varphi^2
\end{equation*}
\begin{equation*}
+2a\sin^2\theta\Big(\frac{2Mr-(Q_e^2+Q_t^2)[W(z)+z(1-z^2)W'(z)]}{r^2+a^2\cos^2\theta}\Big)dtd\varphi,
\end{equation*}
where $z$ is given in~(\ref{Qe_Qc_relation}) 
and depends on the electric and magnetic charges ratio.
Despite on cumbersome structure of the expression~(\ref{KN_line_element_CNED})
it  differs from the line element of Kerr-Newman space-time 
mostly by replacing the term $Q_e^2+Q_t^2$ to ${\cal K}$ 
and is given here only for completeness of description. 
The proximity between the descriptions of the Kerr-Newman 
space-time in the Maxwell theory and in CNED allows to use the 
results obtained in~\cite{a43,a44,a45,a46} 
for the black-hole electromagnetic field, for which it is convenient to use the representation it terms of Faraday two-form:
\begin{equation}\label{F_2form}
F=\frac{1}{2}F_{ik}dx^i\wedge dx^k=\frac{1}{(r^2+a^2\sin^2\theta)^2}
\Big\{\Big[Q_e(r^2-a^2\cos^2\theta)-2aQ_tr\cos\theta\Big]dt \wedge dr
\end{equation}
\begin{equation*} 
-a\sin\theta\Big[2aQ_e r\cos\theta+Q_t(r^2-a^2\cos^2\theta)\Big]dt\wedge d\theta+
a\sin^2\theta\Big[Q_e(r^2-a^2\cos^2\theta)-2aQ_t r\cos\theta\Big]dr \wedge d\varphi
\end{equation*}
\begin{equation*} 
-(r^2+a^2)\sin\theta\Big[2aQ_e r\cos\theta+Q_t(r^2-a^2\cos^2\theta)\Big]d\theta \wedge d\varphi \Big\}.
\end{equation*}
In the asymptotically flat case when $a\ll r$, it is becomes possible to interpret the expression~(\ref{F_2form}) in terms of the multipole expansion:
\begin{equation}
F=E_r \ dt\wedge dr + rE_\theta \ dt\wedge d\theta +r\sin\theta B_\theta \ dr\wedge d\varphi
-r^2\sin\theta B_r\  d\theta\wedge d\varphi,
\end{equation}
where the components of the electric and magnetic field are 
superposition of the monopole therms and dipole 
therms with the electric moment $|{\bf d}|=aQ_t$ produced by 
rotation of the magnetic charge, and the magnetic moment 
$|{\bf m}|=aQ_e$ coupled to the electric charge:
\begin{equation}
E_r=\frac{Q_e}{r^2}-\frac{2aQ_t\cos\theta}{r^3}, 
\qquad E_\theta=-\frac{aQ_t\sin\theta}{r^2},
\end{equation} 
\begin{equation*}
\qquad B_r=\frac{Q_t}{r^2}+
\frac{2aQ_e\cos\theta}{r^3}, 
\qquad 
B_\theta=\frac{aQ_e\sin\theta}{r^3}.
\end{equation*}
As the black hole possess the magnetic dipole moment it will be interesting 
to obtain the gyromagnetic ratio $g$, the definition 
of which in CNED case becomes not entirely unambiguous.
The most commonly  $g$ 
is defined as the multiplier 
in the expression for the magnetic dipole moment:
\begin{equation}\label{g_2definition}
|{\bf m}|=g \frac{Q|{\bf J}|}{2M},
\end{equation}
where $|{\bf J}|=aM$ is a black hole angular 
momentum, $Q$ and $M$ are the electric charge and the
mass respectively. 
Using by the expression for $|{\bf m}|$ 
obtained earlier, it is easy to derive that
$g=2Q_e/Q$, however this raises the question of what type of 
the electric charge $Q$ should be used in~(\ref{g_2definition}).
When $Q=Q_e$ the gyromagnetic ratio is exactly equal to $g_e=2$, as it was obtained in~\cite{a45} even in presence 
of the Lambda-term.
However, if $Q=Q_c$, then in accordance to~(\ref{Qe_Qc_relation}),
the ratio will depend on the choice of CNED model:
\begin{equation}
g_c=2\times \Big\{W(z)+\frac{z(1-z^2)}{2}\Big[1+
\frac{1-z^2}{1+\mbox{sgn}(z)\sqrt{1-(1-z^2)^2}}\Big]W'(z)\Big\}^{-1}.
\end{equation}
The exception is only a special case of CNED 
with the model function $W$ from~(\ref{CNED_sp_W}),  
for which $g_c = g_e = 2$. 
The example above  indicates the possibility of an alternative in definition of the gyromagnetic ratio for CNED, which may be noted  in experimental data  processing.

\section{Conclussion}\label{S7}
In this paper we have considered the main properties of the 
exact classical solutions  describing compact astrophysical 
objects in Einstein's  general relativity and conformal vacuum 
non-linear electrodynamics with the most common form of the 
Lagrangian~(\ref{CNED_L}).
This type of electrodynamics have a set of distinctive features, 
such as traceless of the stress-energy tensor, correspondence to 
all group symmetries of Maxwell theory and the lack of dimensional 
parameter coupled to nonlinearity. To distinguish physically consistent 
models  the restrictions on the Lagrangian  in form of unitarity 
and causality criteria as well as  dominant energy condition were imposed.
It was shown that  an arbitrary  CNED model 
admit a solution in the form of Reissner-Nordstr\"{o}m metric in 
presence of electric and magnetic topological charges, which generalize 
the results obtained earlier by the other authors~\cite{a32}.
To write the Reissner-Nordstr\"{o}m solution, we require two constants with the dimension 
of the electric charge, which can be expressed through each other by 
the relation dependent on choice of the CNED Lagrangian.
One of these constants is a factor in the expression of the 
electric field strength, so it can be called the "force-charge", while 
the other one is a multiplier in the charge density, 
so it was called the "source-charge". 
In the general case, these two charges are different, however 
they will be identical in  the particular case of CNED with 
the model Lagrangian~(\ref{CNED_sp_W}), which is correspondent to
Maxwell theory when the nonlinearity scale parameter is small.

The another one generalization to an arbitrary CNED model was obtained 
for Vaidya-Bonnor solution which describes emission of charged 
null fluid from the point-like center. In the obtained solution, the null 
fluid current density coupled to the central charge variation  by 
the more complicated relation, involving the  Lagrangian model 
function of CNED and it's first two derivatives. Wherein, there is no
any restrictions on the second derivative of the model function 
following from the fundamental principles, so the physical 
self-consistency of the solution should be verified 
in each particular choice of the model Lagrangian.  

Finally, by using Janis-Newman algorithm,
CNED generalization for a Kerr-Newman black hole was obtained.
The analysis of the gyromagnetic ratio for such a black hole 
pointed on a possible ambiguity in the g-factor definition, 
which results from the alternative in the choice of 
the electric charge constant in the
conformal non-linear electrodynamics.

\bibliography{bibl_Conformal}

\end{document}